\documentclass[aps, pra, showpacs, amsmath, twocolumn,
groupedaddress, superscriptaddress]{revtex4}

\usepackage{graphics}

\usepackage{graphicx}

\usepackage{subfigure}

\usepackage{float}

\usepackage{longtable}

\usepackage{amssymb}

\usepackage{mathrsfs}

\usepackage{bm}

\usepackage{natbib}

\begin{document}

\title{High-harmonic generation from arbitrarily oriented diatomic molecules \\
                including nuclear motion and field-free alignment}

\author{C. B. Madsen}
\author{L. B. Madsen}

\affiliation{%
Department of Physics and Astronomy, University of Aarhus, 8000
{\AA}rhus C, Denmark}

\begin{abstract}
We present a theoretical model of high-harmonic generation from
diatomic molecules. The theory includes effects of alignment as well
as nuclear motion and is used to predict results for N$_2$, O$_2$,
H$_2$ and D$_2$. The results show that the alignment dependence of
high-harmonics is governed by the symmetry of the highest occupied
molecular orbital and that the inclusion of the nuclear motion in the 
theoretical description generally reduces the intensity of the harmonic 
radiation. We compare our model with experimental results on N$_2$ and 
O$_2$, and obtain very good agreement.
\end{abstract}


\maketitle

\section{Introduction}\label{Intro}
High-harmonic generation (HHG) is a process in strong-field physics
that occurs when the non-perturbative driving of an electron by an
external field makes it leave its parent atom or molecule, propagate
in the field and finally when the electron is steered back to its
origin, recombine under the emission of a high-energy
photon~\cite{Corkum93}. The process is coherent, so the result of
the interaction is the production of high-frequency coherent
radiation. The emitted signal can be used to create coherent UV or
XUV pulses, and this radiation may then, e.g., in fundamental
science be utilized to probe ultrafast attosecond electron dynamics
(see Refs.~\cite{agostini1,Scrinzi06} for recent reviews).

As compared to atoms, diatomic molecules possess additional degrees
of freedom, namely the nuclear separation and the orientation with
respect to the laser field. Studies of HHG from the H$_2^+$
molecular ion~\cite{lein4,telnov1,bandrauk2} along with the
molecules N$_2$ and O$_2$~\cite{zhou1,lein1} show that the extra
parameters of the molecules give rise to additional effects on the
harmonic spectrum. There are two reasons why this complex behavior
for molecules is of interest: (i) the extra parameters available for
manipulating HHG from the molecules may be useful tools for
controlling the intensity of the harmonics. (ii) the fact that the
harmonic spectrum contains features reminiscent of the molecular
structure means that the emitted signals provide a way to study the
nuclear motion and attosecond electron dynamics of the source of the
high-harmonic radiation~\cite{itani1,lein2,kanai1,itani2}.

The complete theoretical modeling of HHG is computationally very
challenging. In {\it ab initio} approaches the time-dependent
Schr\"{o}dinger equation has to be propagated in a box much larger than is
the case for strong-field ionization studies since in the former
case one has to make sure that the electronic wave packet does not
engage with an absorbing boundary. Additional complications come
from the subsequent propagation of Maxwell's equations in the
generating medium, and in any case going beyond an effective
one-electron model, or other reduced dimensionality models, seems
impossible in the foreseeable future. These complications underline
the importance of developing accurate approximate theories. In
particular for molecules, such models, which should be relatively
inexpensive to evaluate, are of interest since investigations
aiming, e.g., at an increased degree of control of the harmonic
production, require a set of evaluations for different
bond-lengths and orientations with respect to the field. In this
paper we present one such theory. We provide a model to calculate
the angular dependence of the harmonic signal from a single diatomic
molecule of arbitrary orientation with respect to the laser field.
In particular we include the effect of nuclear motion (vibration),
and we test the accuracy of the model by comparison with experiments.
High-harmonic generation from diatomic molecules has
already been considered in an extension of the Lewenstein
model~\cite{Lewenstein94} in Ref.~\cite{zhou1}. In that work
depletion from the initial state was considered but the nuclear
motion was ignored. Recently, the strong-field approximation
(SFA) for HHG in diatomic molecules subject to ultrashort laser
pulses was presented, again ignoring nuclear motion, but combined with a
careful saddle-point analysis and a discussion of the gauge
problem~\cite{Chirila06}. Here, we are concerned with the response
to a monochromatic field, and in parallel with the conventions in
the theory of strong-field ionization, we will refer to our theory
as the length gauge version of the molecular SFA for HHG.

The present paper is organized as follows. In Sec.~\ref{Theory} we
present the theory for HHG from diatomic molecules in monochromatic
linearly polarized laser fields. Section~\ref{CalcDet} contains some
calculational details. In Sec.~\ref{sectionResults} we show the HHG
yields for N$_2$ and O$_2$ as a function of the orientation of the
molecules with respect to the laser polarization, and compare with
experimental measurements. By discussing HHG signals from the
isotopes H$_2$ and D$_2$, we also see how a difference in nuclear
mass affects the harmonic spectrum. The paper ends with a
summary and a conclusion in Sec.~\ref{Conclusion}.
Appendix~\ref{AppendixA} contains a discussion of the momentum space
wave function of a molecule in the single-active-electron (SAE)
approximation.

\section{Theory}\label{Theory}
 In our treatment of the HHG from molecules, we
extend the atomic model of Ref.~\cite{ostrovsky1}, and include
notions of nuclear motion. We consider the case, where the laser pulse
contains several cycles such that a Floquet approach is suitable.
As we shall see below when we compare with experimental results,
even for probe pulses of 50 fs~\cite{kanai1} and 35 fs~\cite{itani2}
duration this turns out to be an accurate approximation. Hence,
using the Coulomb gauge and the dipole approximation, a laser field
with frequency $\omega$ and period $T=2\pi/\omega$ is described by
the vector potential $\mathbf{A}(t)=\mathbf{A}_0\cos(\omega t)$. The
electric field, $\mathbf{F}(t)$, is then
$-\partial\mathbf{A}(t)/\partial t$, i.e.,
$\mathbf{F}(t)=\mathbf{F}_0\sin(\omega t)$ with
$\mathbf{F}_0=\omega\mathbf{A}_0$ [atomic units,
$e=\hbar=m_e=a_0=1$, are used throughout unless indicated
otherwise].

The harmonic generation is determined by the dipole moment
\begin{equation}\label{exactdipole}
  d(t)=\left<\Psi(t)\right| \hat{D}_{\boldsymbol{\epsilon}}^{(n)}
  \left|\Psi(t) \right>,
\end{equation}
where the dipole operator is defined by
\begin{equation}
  \hat{D}_{\boldsymbol{\epsilon}}^{(n)} =
  \sum_{i=1}^n\hat{d}_{\boldsymbol{\epsilon},i}
  =\sum_{i=1}^n\boldsymbol{\epsilon}\cdot\mathbf{r}_i,
\end{equation}
with $\boldsymbol{\epsilon}$ the polarization vector of the field,
and where $| \Psi(t) \rangle$ is the solution to the $n$-electron
Schr{\"o}dinger equation for the diatomic molecule
\begin{equation}\label{schroedinger}
  \left[ i\frac{\partial}{\partial t} - H_0-V_F^{(n)} \right]
  \Psi\left(\mathbf{r}_1,\mathbf{r}_2,\ldots,\mathbf{r}_n,
  \mathbf{R},t \right)= 0,
\end{equation}
with $\mathbf{r}_i$ the coordinate of electron $i$, and $\mathbf{R}$
the relative position of the two nuclei, $H_0$ the field-free part
of the Hamiltonian of the molecule, and $V_F^{(n)}$ the interaction of the
molecule with the laser field. In the present length gauge description
\begin{equation}\label{laser-molecule-interaction}
V_F^{(n)}=\mathbf{F}(t)\cdot\sum_{i=1}^n\mathbf{r}_i.
\end{equation}
From Eq.~\eqref{exactdipole} it is clear, that in order to obtain
the dipole moment, we need a solution of Eq.~\eqref{schroedinger}.
The state $\left| \Psi(t) \right>$  is determined by using the
description of the laser-molecule system given in
Ref.~\cite{kjeldsen1}. Here nuclear motion is modelled within the
Born-Oppenheimer (BO) approximation, and, assuming negligible
electron-electron correlation, the electrons are put into orbitals
obtained from Hartree-Fock (HF) calculations. Consequently, the
interaction of the molecule can be treated in the SAE approximation,
where the active electron is the electron in the highest occupied
molecular orbital (HOMO). We write $\left| \Psi(t) \right>$ as a
superposition of the field-free state of the molecule and final
states, where the active electron is excited to the continuum.
The expansion coefficients are obtained under the assumption
that there is only limited ionization, which again
requires that the laser intensity is below saturation (the regime of 
relevance in the HHG process). Inserting the expanded state into 
Eq.~\eqref{exactdipole}, we arrive at the following approximation
to the dipole moment
\begin{widetext}
\begin{eqnarray}\label{dipolmoment}
  d_0(t) = -i\int_{-\infty}^{t} dt' \sum_{\nu_i}\int d\mathbf{q}
  \left< \Phi_0(t)\right|\hat{D}_{\boldsymbol{\epsilon}}^{(n)}\left|
  \Psi_{\nu_i,\mathbf{q}}(t)\right> \left< \Psi_{\nu_i,\mathbf{q}}(t')
  \right|V_F^{(n)}(t') \left| \Phi_0(t') \right>.
\end{eqnarray}
\end{widetext}
Here
\begin{eqnarray}\label{field-free-groundstate}
  \Phi_0 &=& \frac{1}{\sqrt{n!}} \det \left[
    \phi_1(\mathbf{r}_1;R_0) \phi_2(\mathbf{r}_2;R_0)\ldots
    \phi_n(\mathbf{r}_n;R_0) \right] \nonumber \\
    &\times&\chi_{\nu_0}(R)\exp(-iE_0 t)
\end{eqnarray}
is the field-free ground state of total energy
$E_0=E_0^e(R_0)+E_{\nu_0}$, where $E_0^e(R_0)$ denotes the
electronic eigenenergy at nuclear equilibrium distance $R_0$, and
$E_{\nu_0}$ is the vibrational eigenenergy of the nuclear
Hamiltonian. The electronic part of this state is expressed as a
Slater determinant of the orbitals $\phi_j(\mathbf{r}_i;R_0)$, while
$\chi_{\nu_0}(R)$ denotes the vibrational ground state wave function,
labeled by vibrational quantum number $\nu_0$. Furthermore
\begin{eqnarray}\label{field-state}
  \Psi_{\nu_i,\mathbf{q}} &=&  \frac{1}{\sqrt{n!}} \det \left[
  \phi_1(\mathbf{r}_1;R_0) \phi_2(\mathbf{r}_2;R_0) \ldots
  \Phi_{\mathbf{q}}(\mathbf{r}_n,t) \right] \nonumber \\
  &\times& \chi_{\nu_i}^+(R)\exp(-iE^+ t).
\end{eqnarray}
is the state of the molecule interacting with the laser field and
$E^+=E^{e,+}(R_0)+E_{\nu_i}^+$ is the total energy of the residual
molecular ion. We also introduced $\Phi_\mathbf{q}$, which
denotes the state of the active electron, and we neglect the
interaction of this electron with the residual molecular ion. Then the
state of the active electron is given by a Volkov wave
\begin{eqnarray}\label{volkovwave}
  \Phi_{\mathbf{q}}(\mathbf{r},t)&=&\frac{1}{(2\pi)^{3/2}}\exp
  \Bigg\{i \Bigg[(\mathbf{q}+\mathbf{A}(t))\cdot
  \mathbf{r} \nonumber \\
  &-&\mathbf{q}\cdot\boldsymbol{\alpha}_0 \sin(\omega
  t)-\frac{U_p}{2\omega}\sin(2\omega t) \nonumber \\
  &-&\left(\frac{q^2}{2}+U_p\right) t \Bigg] \Bigg\},
\end{eqnarray}
with quiver radius
$\boldsymbol{\alpha}_0=\mathbf{A}_0/\omega$ and ponderomotive
potential $U_p=A_0^2/4$.

Using the Slater-Condon rules~\cite{condon1}, the expression in
Eq.~\eqref{dipolmoment} reduces to an expression involving only 
the electron coordinate of the active electron
\begin{widetext}
\begin{eqnarray}\label{simple-dipole-moment}
  d_0(t) &=& -i \sum_{\nu_i}\int_{-\infty}^{t} dt' \int
  d\mathbf{q} \left<\nu_0|\nu_i\right> \left<\phi_0(\mathbf{r};
  R_0)\right|\hat{d}_{\boldsymbol{\epsilon}}\left|
  \Phi_{\mathbf{q}}(\mathbf{r},t)\right> \exp \left[i
  \left( E_0^e(R_0)+E_{\nu_0} -E^{e,+}(R_0)-E_{\nu_i}^+
  \right)t\right] \nonumber \\
  &\times& \left<\nu_i|\nu_0\right>
  \left<\Phi_{\mathbf{q}}(\mathbf{r}',t')\right| V_F(t')
  \left|\phi_0(\mathbf{r}';R_0) \right>
  \exp\left[-i\left(E_0^e(R_0)+E_{\nu_0}-E^{e,+}(R_0)-
  E_{\nu_i}^+\right)t'\right].
\end{eqnarray}
\end{widetext}
To simplify notation, we have dropped the index of the electron and 
consequently introduced 
$\hat{d}_{\boldsymbol{\epsilon}}=\boldsymbol{\epsilon}\cdot\mathbf{r}$,
 $V_F(t')=\mathbf{F}(t')\cdot\mathbf{r}'$. Also we denote the HOMO 
 wave function by $\phi_0(\mathbf{r})$, and
\begin{equation}
 \left<\nu_i|\nu_0\right> = \int_0^\infty dR \chi_{\nu_i}^{+*}(R) \chi_{\nu_0}(R)
\end{equation}
denotes the Franck-Condon (FC) factor.
\subsection{The harmonics and the three-step picture}
The harmonic spectrum is calculated from the Fourier transform of
the dipole moment $d_0(t)$, which implies that the intensity of the
$N$th harmonic is proportional to the norm square of the $N$th Fourier
component, $d_N$. This Fourier component is investigated
by extending the technique of Ref.~\cite{ostrovsky1} to the
molecular case
\begin{widetext}
\begin{eqnarray}\label{dN}
  d_N &=&  \frac{1}{T}\int_0^T dt\exp(iN\omega t)d_0(t)= -\sum_k \sum_{\nu_i} \frac{1}{T}
  \int_0^T dt\frac{1}{T}\int_{0}^{T} dt' \nonumber \\
  &\times& (2\pi)^2 \left<\nu_0|\nu_i\right>
  \left<\right.\phi_0(\mathbf{r})|\exp(iN\omega
  t)\hat{d}_{\boldsymbol{\epsilon}}|
  \Phi_{\mathbf{K}_k^{\nu_i}}(\mathbf{r},t)\left.\right> \exp\left[ i\left(
  E_0^e(R_0)+E_{\nu_0}-E^{e,+}(R_0)-E_{\nu_i}^+\right) t\right]\frac{1}{L_0(t,t')}\nonumber \\
  &\times& \left<\right.\nu_i|\nu_0\left.\right> \left<\right.
  \Phi_{\mathbf{K}_k^{\nu_i}}(\mathbf{r}',t')| V_F(t')
  \phi_0(\mathbf{r}') \left.\right> \exp\left[
  i\left(E^{e,+}(R_0)+E_{\nu_i}^+-E_0^e(R_0)-E_{\nu_0}\right) t'\right].
\end{eqnarray}
\end{widetext}
An interpretation of Eq.~\eqref{dN} is possible in terms of the three-step
picture of HHG. The matrix element $\left<\right.\Phi_{\mathbf{K}_k^{\nu_i}}(\mathbf{r}',t')
| V_F(t') |\phi_0(\mathbf{r}') \left.\right>$ describes the transition from the 
HOMO and into the field-dressed Volkov state with momentum of magnitude
\begin{equation}\label{volkovwave-momentum}
 K_k^{\nu_i} = \sqrt{2(k\omega+E_0^e(R_0)+E_{\nu_0}-E^{e,+}(R_0)-E_{\nu_i}^+-U_p)}.
\end{equation}
The direction of the momentum is parallel or anti-parallel (prescribed by the 
positive or negative sign of $\sigma$ defined below) to the polarization vector 
of the laser field. This momentum emerges after performing the $\mathbf{q}$ 
integration in Eq.~\eqref{simple-dipole-moment}, provided
that the laser field is sufficiently strong 
in the sense that the quiver radius, $\boldsymbol{\alpha}_0$, 
is large compared to the extent of the molecule. Following~\cite{ostrovsky1}, we 
include only terms corresponding to real values of $K_k^{\nu_i}$, because of 
the clear physical interpretation of Eq.~\eqref{dN}. The exponential of the last 
line of Eq.~\eqref{dN} describes the energy balance in this initial step, and 
$\left<\nu_i|\nu_0\right>$ is the FC factor associated with the transition in 
the nuclear degree of freedom. The factor $1/L_0(t,t')$ with 
$L_0(t,t')=\sigma\alpha_0(\sin\omega t'-\sin\omega t)$ and $\sigma=\pm 1$ to assure 
$\mathrm{Re}(L_0)>0$ comes from the Volkov propagator and describes the decrease of 
the amplitude of the continuum wave as the latter expands into three-dimensional space. 
Finally, the matrix element 
$\left<\right.\phi_0(\mathbf{r})|\hat{d}_{\boldsymbol{\epsilon}}
|\Phi_{\mathbf{K}_k^{\nu_i}}(\mathbf{r},t)\left.\right>$ with associated phase 
factors and FC factor describes the laser-assisted recombination (LAR) step
of the three-step model. We note that, in our model, the
expression of Eq.~\eqref{dN} predicts a correlation of the momentum
of the continuum electron with the nuclear motion of the residual
molecular ion, through the oscillating time-dependent exponential factors.

\subsection{The harmonic spectrum from arbitrarily oriented
molecules} In order to be able to evaluate Eq.~\eqref{dN} for
molecules of arbitrary orientation with respect to the laser field
we use a HOMO wave function, where the angular part is
expanded onto spherical harmonics. In the body-fixed frame, where
the $z$ axis is chosen along the internuclear axis, the HOMO wave
function is then given by
\begin{equation} \label{homoex}
  \phi_0^{bf}(\mathbf{r})=\sum_{l} F_{l,m}(r)
  Y_l^m(\hat{\mathbf{r}}).
\end{equation}
We also need the asymptotic form of this expression
\begin{equation} \label{atiwave}
  \phi_0^{bf}(\mathbf{r})\sim \sum_{l} C_{l,m} r^{Z / \kappa -1}
  \exp(-\kappa r) Y_l^m(\hat{\mathbf{r}}).
\end{equation}
Here $\kappa = \sqrt{-2E_b}$, $E_b$ is the binding energy of 
the active electron in the HF ground state wave function and $Z$ 
is the net charge of the molecule, when the HOMO electron is removed. 
The $t'$ integral of Eq.~\eqref{dN} corresponds to the ATI step. It is well 
known that in this process the HOMO electron escapes to the laser-dressed
continuum, when the distance to the nucleus is
large~\cite{ostrovsky2}. As a result we can employ the asymptotic form of
the HOMO in the $t'$ integrand in Eq.~\eqref{dN}, and the integral
will subsequently be evaluated using the saddle-point method. This
evaluation proceeds as in Ref.~\cite{kuchiev1}, and the HOMO as given by 
Eq.~\eqref{atiwave} is replaced in favor of the momentum space wave function.
It is sufficient to know the latter near its singular points, where it is 
given by the particularly simple expression Eq.~\eqref{singularity-wave}, as 
described in Appendix A.
The $t$ integration occurring in Eq.~\eqref{dN} corresponds to the LAR
step and is performed numerically. Previous publications show that in
the LAR step it is inadequate to approximate the HOMO with the asymptotic
form, because the recombination takes place at small separations of
the excited electron and the molecular core~\cite{bandrauk2,ostrovsky4}.
Also here we prefer to replace the space wave function by from Eq.~\eqref{homoex}
by the momentum space wave function, which in the body fixed frame may be written 
in the following way
\begin{equation} \label{larwave}
 \tilde{\phi}_{0}^{bf}(\mathbf{q})=\sum_{l} G_{l,m}(q)
 Y_l^m(\hat{\mathbf{q}}),
\end{equation}
where the $G_{l,m}$'s are obtained numerically. Because the HOMO wave 
functions are expressed in terms of spherical harmonics, we can easily 
obtain the wave functions of the molecule in the laboratory frame, 
where the internuclear axis does not necessarily coincide with the $z$ 
axis. For a molecule, where the internuclear axis is rotated with respect 
to the $z$ axis of the laboratory frame, the rotation is characterized 
by the three Euler angles 
$\mathbf{\mathscr{R}}=(\alpha,\beta,\gamma$)~\cite{sakurai}, and we
arrive at the following expression of the $N$th harmonic for a molecule of
orientation $\mathbf{\mathscr{R}}$ with respect to the laser polarization
\begin{widetext}
\begin{eqnarray} \label{harmonic}
  d_N =
  \sum_{l_2,l_1}\sum_{m_2',m_1'}D_{m_2',m_2}^{l_2*}
  (\mathbf{\mathscr{R}})D_{m_1',m_1}^{l_1}
  (\mathbf{\mathscr{R}})C_{l_1,m_1}\sum_{\nu_i}|\left<\nu_i|\nu_0\right>|^2
  \sum_{k}\sum_{sp(k)}\mathscr{B}
  _{l_2,m_2',m_2}^{N,\nu_i,k}(sp(k))\mathscr{A}_{l_1,m_1'}^{\nu_i,k}
  (sp(k)),
\end{eqnarray}
with
\begin{eqnarray}\label{Aamp}
  \mathscr{A}_{l_1,m_1'}^{\nu_i,k}(sp(k)) = -\frac{1}{T}\Gamma
  \left(1+\frac{Z / \kappa}{2}\right) 2^{\frac{Z / \kappa}{2}}
  \kappa^{Z / \kappa}(\pm 1)^{l_1}\frac{\exp [iS(t_{sp(k)}')]}
  {\sqrt{[-iS''(t_{sp(k)}')]^{1+Z / \kappa}}}
  \left. Y_{l_1}^{m_1'}\left(\hat{\mathbf{q}}'\right)\right
  \arrowvert_{\mathbf{q}'=\mathbf{K}_k^{\nu_i}+\mathbf{A}(t_{sp(k)}')},
\end{eqnarray}
and
\begin{eqnarray}\label{Bamp}
  \mathscr{B}_{l_2,m_2',m_2}^{N,\nu_i,k}(sp(k)) = i \frac{(2\pi)^2}
  { T} \int_0^T dt \frac{\exp[i(N\omega t-S(t))]}{L_0(t,t_{sp(k)}')}
  (\boldsymbol{\epsilon}\cdot \mathbf{\nabla}_
  {\mathbf{q}} ) \left. \left[ G_{l_2,m_2}(q)Y_{l_2}^{m_2'}
  \left(\hat{\mathbf{q}}\right) \right]^*\right\arrowvert_{
  \mathbf{q}=\mathbf{K}_k^{\nu_i}+\mathbf{A}(t)}.
\end{eqnarray}
\end{widetext}
Here $D_{m_i',m_i}^{l}(\mathbf{\mathscr{R}})$, $i=1,2$ denotes the
matrix elements of the rotation operator~\cite{sakurai} and
\begin{equation}\label{action}
  S(t) = k\omega t+\mathbf{K}_k^{\nu_i}\cdot \boldsymbol{\alpha}_0
  \sin(\omega t) + \frac{U_p}{2\omega}\sin(2\omega t)
\end{equation}
is the quasi-classical action. The index $sp(k)$ denotes the saddle-points. 
For each $k$ the saddle-points $t'_{sp}(k)$ are defined by the condition
$S'(t'_{sp}(k))=0$, and we use the ones with $0 \leq
\mathrm{Re}(t_{sp}'(k))< T$ along with $\mathrm{Im}(t_{sp}'(k))>0$.
Note that Eq.~\eqref{Aamp} is evaluated in the complex $t'$ plane and the 
factor $(\pm 1)^{l_1}$ in Eq.~\eqref{Aamp} corresponds to the
limits $\pm i\kappa$ of the size $q'$ of the electron momentum at the
saddle-points. Letting the laser polarization define the $z$ axis of
the laboratory frame, we have both the momentum
of the Volkov wave as well as the vector potential parallel to the $z$ axis, so that $Y_{l_1}^{m_1'}(\hat{\mathbf{q}}')\propto\delta_{m_1',0}$
in Eq~\eqref{Aamp}. Equation~\eqref{Bamp} is evaluated along the real $t$ axis, and 
when the polarization vector of the field is directed along the $z$ axis 
$\boldsymbol{\epsilon}\cdot \mathbf{\nabla}_{\mathbf{q}}$ reduces to a partial derivative 
with respect to the size $q$ of the momentum. As mentioned earlier we consider only 
values of $k$, where $K_k^{\nu_i}$ is real, which serves to define a lower limit $k_{min}$
of the sum. The upper limit $k_{max}$ is determined by requiring
convergence of the calculated value of $d_N$. Equation~\eqref{harmonic}
along with the accompanying definitions of Eqs.~\eqref{Aamp}-\eqref{action}
summarize the formulation of our model. A cruder estimate of the harmonic 
generation can be obtained by using the asymptotic form of the HOMO 
wave function in both the ATI and the LAR step. In this approximation the 
$N$th harmonic is given by
\begin{widetext}
\begin{eqnarray} \label{harmonicas}
  d_N^{as} &=&
  \sum_{l_2,l_1}\sum_{m_2',m_1'}D_{m_2',m_2}^{l_2*}
  (\mathbf{\mathscr{R}})D_{m_1',m_1}^{l_1}
  (\mathbf{\mathscr{R}})C_{l_2,m_2}^* C_{l_1,m_1}
  \sum_{\nu_i}|\left<\nu_i|\nu_0\right>|^2\sum_{k}\sum_{sp(k)}\tilde{\mathscr{B}}
  _{l_2,m_2'}^{N,\nu_i,k}(sp(k))\mathscr{A}_{l_1,m_1'}^{\nu_i,k}(sp(k)),
\end{eqnarray}
with
\begin{eqnarray}\label{eq: Bampas}
  \tilde{\mathscr{B}}_{l_2,m_2'}^{N,\nu_i,k}(sp(k)) &=& - \frac{\sqrt{2}
  (2\pi)^2}{ T} \int_0^T dt \frac{\exp[i(N\omega t-S(t))]}{L_0(t,t_{sp}'(k))}
  (-i\boldsymbol{\epsilon}\cdot \mathbf{\nabla}_{\mathbf{q}} )
  \left[ \left(\frac{q}{i\kappa}\right)^{l_2}
  \frac{1}{2^{l_2+1}\kappa^{2+Z / \kappa}}\frac{\Gamma (l_2+2+Z / \kappa)}{\Gamma(l_2+
  \frac{3}{2})}Y_{l_2}^{m_2'}\left(\hat{\mathbf{q}}\right)  \right.
  \nonumber \\
  &\times& \left. \left._2F_1\left(\frac{l_2+2+Z / \kappa}{2},\frac{l_2+2+Z / \kappa}{2}
  + \frac{1}{2}; l_2+\frac{3}{2}; -\left(\frac{q}{\kappa}\right)^2\right) \right]^*\right\arrowvert_{ \mathbf{q}=\mathbf{K}_k^{\nu_i}+\mathbf{A}(t)},
\end{eqnarray}
\end{widetext}
where all functions are analytically known and where the
$C_{lm}$ parameters are available in Table I (see also Table I in
Ref.~\cite{kjeldsen1}). The asymptotic expression in
Eq.~\eqref{harmonicas} has the advantage that all geometric
factors are pulled outside the integration.

In Sec. \ref{sectionResults} we compare HHG spectra obtained
with the two forms of Eqs.~\eqref{harmonic} and~\eqref{harmonicas}.
In that section we also show curves, which do not include nuclear
motion. When the nuclei are clamped, the harmonic strengths are
simply given by including only the first term in the sum over
$\nu_i$ in Eqs.~\eqref{harmonic} and~\eqref{harmonicas} and
replacing the Franck-Condon factors by unity.

\section{Calculational details}\label{CalcDet}
In this section we outline some calculational details to be used
below. First we sketch the procedure used to obtain the
$G_{l,m}(q)$'s and $C_{l,m}$'s of Eqs.~\eqref{harmonic}
and~\eqref{harmonicas}. The initial step in the procedure is to solve
the HF equations fully numerically for the diatomic
molecules~\cite{Kobus96}. In this manner the ground state orbitals
are obtained, and are projected onto spherical
harmonics to attain the radial functions $F_{l,m}(r)$. The
asymptotic angular coefficients are found by comparing the HF
radial wave functions with the form $C_{l,m}r^{Z / \kappa-1}\exp(-\kappa
r)$, where the $C_{l,m}$'s are used as fitting parameters. The
$G_{l,m}(q)$'s, on the other hand, are obtained from the Fourier-Bessel
transform of the $F_{l,m}(r)$'s.

In the cases of N$_2$ and O$_2$ we compare results of our model with
experiments where the degree of molecular alignment was controlled
by non-adiabatically creating a rotational wavepacket by a pump
pulse and then monitoring the harmonic yield as a function of the
time delay to the intense probe pulse~\cite{itani2,kanai1}. In order
to perform a comparison with the experimental results, we need the
time-dependent angular distribution $\rho(t,\beta)$ of the
rotational wave packet created by the pump pulse. We refer to
Ref.~\cite{ortigoso} for details about calculations of rotational
wave packets. We assume that  rotation is negligible under the probe
pulse, and simply note at this point that once the angular
distribution is available, the amplitude of the $N$th harmonic at time
$t$ is given by
\begin{equation}\label{eq:dNtime}
\bar{d}_N(t)=\int_0^\pi d\beta \sin(\beta) \rho(t,\beta)d_N(\beta).
\end{equation}
where $d_N(\beta)$ can be found using Eq.~\eqref{harmonic} for a probe
laser parallel to the pump laser. We can then calculate the
intensity of each harmonic from the rotational wave packet at every
point of time.

Table \ref{tab:table1} shows relevant information for the
calculations presented in the next section.
\begin{table*}
\caption{\label{tab:table1}The molecular properties used in this
work for evaluation of HHG. $I_p$ is the experimental adiabatic
ionization potential, $R_0$ is the equilibrium distance and $B$ is
the rotational constant of the molecule~\cite{NIST}. The parallel
and perpendicular polarizabilities $\alpha_{\parallel}$ and
$\alpha_{\perp}$ of N$_2$ and O$_2$ are based on values
from~\cite{torres} and~\cite{NIST}, respectively. We furthermore give
the angular coefficients $C_{l,m}$ from HF calculations. FC factors and
vibrational energies can be found in the references indicated after
each molecular species.}
\begin{ruledtabular}
\begin{tabular}{cccccccccc}
 & &$I_p$ (eV)&$R_0$ (\AA)& $B$ (GHz) & $\alpha_{\parallel}$ (\AA$^3$) & $\alpha_{\perp}$ (\AA$^3$) & $C_{00}$ & $C_{2m}$ &$C_{4m}$\\
\hline
N$_2$ $(\sigma_g)$& \cite{halmann1}                & 15.58  & 1.098 & 59.647 & 2.38 & 1.45 & 3.46 & 1.64 & 0.12 \\
O$_2$ $(\pi_g)$   & \cite{nicholls1}               & 12.03  & 1.208 & 42.861 & 2.3  & 1.1  &      & 1.04 & 0.07  \\
H$_2$ $(\sigma_g)$& \cite{gordon1}\footnotemark[1] & 15.43  & 0.741 &        &      &      & 2.44 & 0.14 &  \\
D$_2$ $(\sigma_g)$& \cite{gordon1}                 & 15.47  & 0.742 &        &      &      & 2.44 & 0.14 &  \\
\end{tabular}
\end{ruledtabular}
\footnotetext[1]{Vibrational energies from Ref.~\cite{cohen1} are
used.}
\end{table*}
\\
\section{Results and discussion}\label{sectionResults}
\subsection{The N$_2$ molecule}
\begin{figure}[ttb]
\includegraphics[width=\columnwidth]{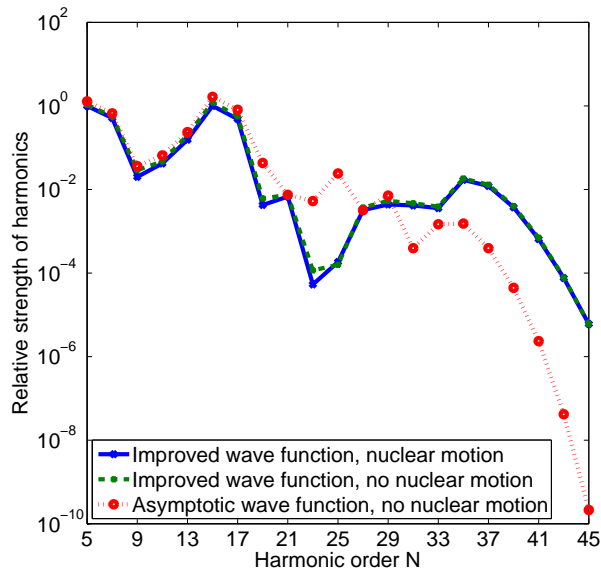}\\
\caption{(color online). Comparison of HHG from the N$_2$ molecule
when using the asymptotic wave function in both the LAR and ATI
step, and when the wave function used in the LAR step is improved.
Only the solid curve includes nuclear motion. The signals have been
normalized to the strength of the fifth harmonic obtained by using
the improved wave function in the LAR step and including nuclear
motion. The internuclear axis and the polarization are parallel
($\beta=0^{\circ}$).}\label{fig:fig3}
\end{figure}

We begin by presenting a calculation of HHG from the N$_2$ molecule
with a HOMO of $\sigma_g$ symmetry. The laser is assumed to have an
energy of $\omega=0.057$ a.u. corresponding to a wavelength of
$\lambda=800$ nm, and the intensity is taken to be
$I=2\times10^{14}$ W/cm$^2$ which is well below the saturation
laser intensity~\cite{itani2}. Such a choice of parameters is within
the range of a Ti:Sapphire laser. Figure~\ref{fig:fig3} shows the
relative strength of the harmonics as a function of the harmonic
order in a geometry, where the internuclear axis of the molecule
is aligned parallel to the polarization of the laser ($\beta=0^{\circ}$).
Note that we refer to the norm square of the Fourier components
(i.e. $|d_N|^2$) as the strength of the harmonics. There
are three different curves on the figure, and the dotted curve
corresponds to Eq.~\eqref{harmonicas}, i.e., the case where the
asymptotic wave function is used in both the ATI and LAR step of the
HHG process, while the dashed and solid curves corresponds to
Eq.~\eqref{harmonic} meaning that the improved wave function was
employed in the LAR step. Calculations represented by the solid
curve take nuclear motion into account.
The harmonic spectrum of the N$_2$ molecule is different, when the
asymptotic wave function is used everywhere as compared to the case,
when the improved wave function is employed in the LAR step. This
result shows that the quality of the HOMO wave function is of
importance. It is also clear from Fig.~\ref{fig:fig3} that the
effect of nuclear motion in the case of N$_2$ is negligible. This is
because the BO potential energy curves of the N$_2$
molecule and the N$_2^+$ molecular ion are nearly parallel, which
means that ionization of an N$_2$ molecule in the ground state
always leaves the residual molecular N$_2^+$ ion in the vibrational
ground state.

\begin{figure}[ttb]
\includegraphics[width=\columnwidth]{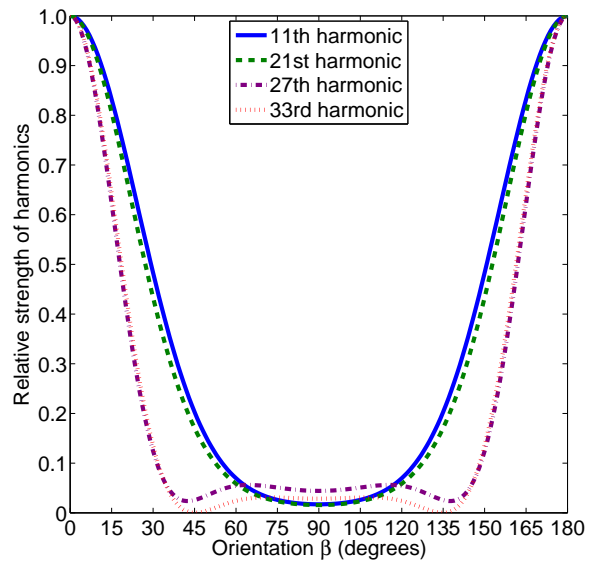}\\
\caption{(color online). Orientation dependence of the harmonics for
the molecule N$_2$. We have employed the improved wave function in
the LAR step and included nuclear motion. All curves are normalized
to their respective values at $\beta = 0^{\circ}$.}\label{fig:fig4}
\end{figure}

In Fig.~\ref{fig:fig4} we study the orientation dependence of the HHG yield from
N$_2$.
It is seen from the figure that harmonic generation for the parallel
alignment ($\beta = 0^{\circ}$) is more intense than the harmonic
generation for the perpendicular alignment ($\beta = 90^{\circ}$),
which reflects the orientation dependence of the ionization of the
molecule (see Ref.~\cite{kjeldsen2} and references therein). The detailed
angular behavior depends on the harmonic order in consideration: a
minimum is observed at approximately $40^{\circ}$ for some orders.
A similar behavior was recently predicted in Ref.~\cite{zhou1}, where
the minima were located at a relative orientation close to $60^{\circ}$.

\begin{figure}[ttb]
\includegraphics[width=\columnwidth]{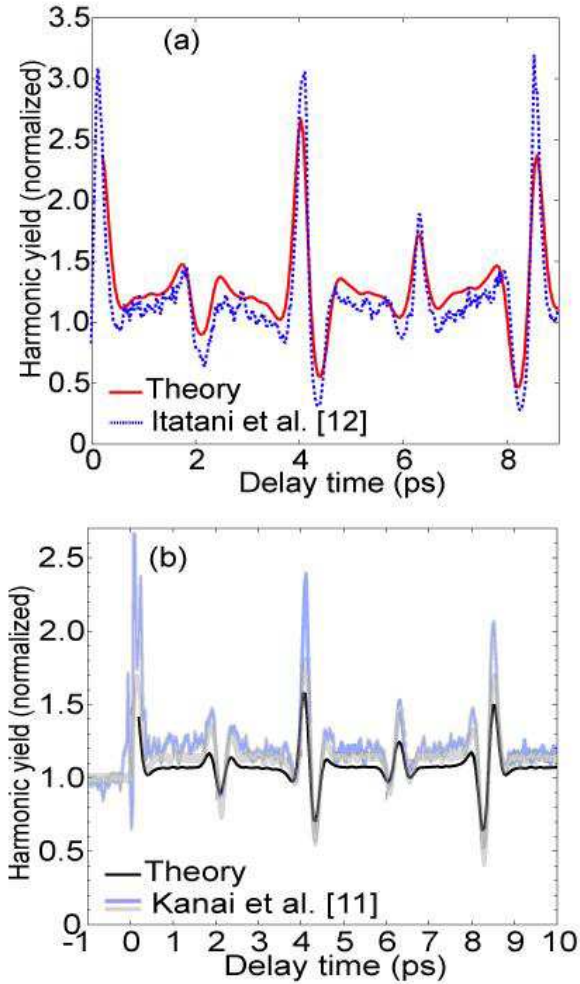}\\
\caption{(color online). Comparison of theory and experiment in the
case of N$_2$. Panel~(a) shows the average of the $21$st to the $25$th 
harmonics from a rotational wave packet of N$_2$. Theoretical results are
based on our model, while the results of Itatani et al. are experimental
data from Ref.~\cite{itani2}. In panel~(b) the $23$rd harmonic according to our 
model is compared with the experimental results of Ref.~\cite{kanai1}.
Our results are normalized to the randomly aligned case.}\label{fig:fig01}
\end{figure}

Several experiments have measured the harmonic emission from
non-adiabatically aligned N$_2$
molecules~\cite{itani2,kanai1,miyazaki}. We can simulate these
results as outlined in Sec.~\ref{CalcDet}. To compare with the
results of Ref.~\cite{itani2} we use the time-dependent angular
distribution $\rho(t,\beta)$ corresponding to a rotational wave
packet created by the experimental pump pulse, $60$ fs, $800$ nm,
intensity $4\times10^{13}$ W/cm$^2$, and the experimental initial
rotational temperature of N$_2$ of $30$ K. To model the results from
Ref.~\cite{kanai1} we obtain the time-dependent angular distribution
with the experimental pump pulse of duration $50$ fs and with
wavelength $800$ nm and intensity $6\times10^{13}$ W/cm$^2$. In
Eq.~\eqref{eq:dNtime} the value of $d_N(\beta)$ is calculated using 
Eq.~\eqref{harmonic} for a laser of wavelength $\lambda=800$ nm and
intensity $I = 2\times10^{14}$ W/cm$^2$, as in Refs.~\cite{itani2,kanai1}.
We can then find the intensity of each harmonic from the rotational
wave packet at every point of time. On Fig.~\ref{fig:fig01} we
compare model calculations with experimental results. The figure shows
the evolution of the harmonic emission from the N$_2$ molecule as a
function of the delay time between the pump and probe pulse. In general
the agreement between experiments and theory is very satisfactory.

\subsection{The O$_2$ molecule}
\begin{figure}[ttb]
\includegraphics[width=\columnwidth]{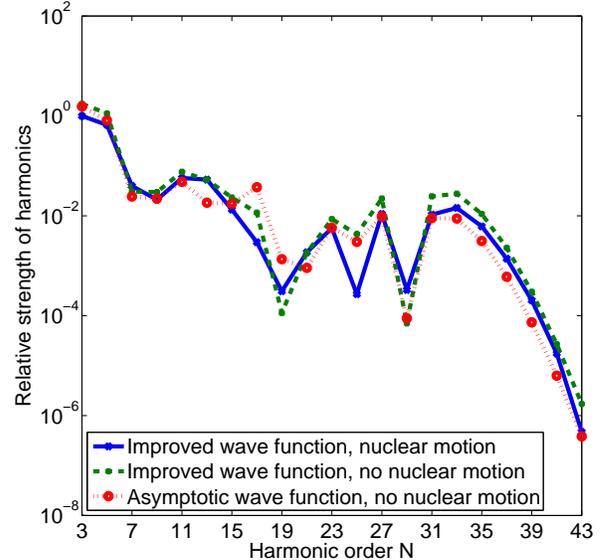}\\
\caption{(color online). HHG from the O$_2$ molecule. The strength
of the harmonics is normalized to the strength of the third harmonic
obtained in the calculation, where an improved wave function is used
in the LAR step and the nuclear motion is included. The internuclear
axis is oriented at an angle $\beta=36^{\circ}$ with respect to the
internuclear axis. See text for further details.}\label{fig:fig8}
\end{figure}

We next consider the O$_2$ molecule, which has a HOMO with $\pi_g$
symmetry. The laser parameters are the same as in the preceding
subsection. Figure \ref{fig:fig8} shows the relative
strength of the harmonics for $\beta=36^{\circ}$.
As in the case of N$_2$ it is important to use a better wave
function than the asymptotic one in the LAR step. The effect of
nuclear motion is more pronounced than for the N$_2$ molecule, because 
the Franck-Condon distribution of O$_2$ is broader than
that of the N$_2$ molecule (see Ref.~\cite{KM-prl05} for more
discussion on the implications on final state vibrational levels and
possible control of these).

\begin{figure}[ttb]
\includegraphics[width=\columnwidth]{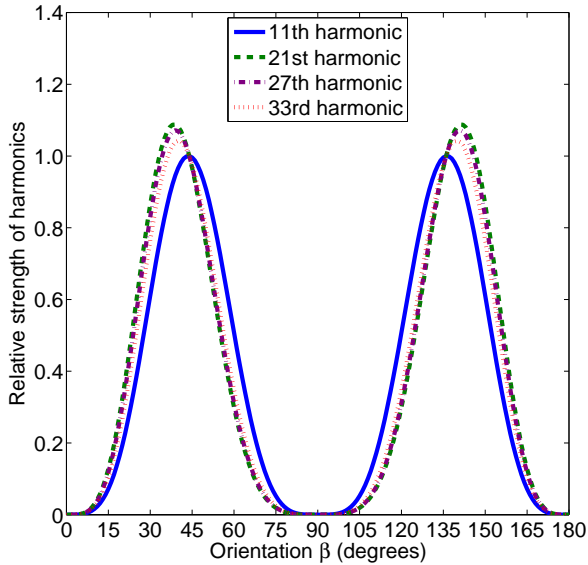}\\
\caption{(color online). HHG from the O$_2$ molecule as a function
of the orientation. The strength of the harmonics are calculated
using Eq.~\eqref{harmonic}. All curves are normalized to their value
at $\beta= 45^{\circ}$.}\label{fig:fig9}
\end{figure}

We now turn to the angular dependence of the harmonics. The strength
of some selected harmonics as a function of the orientation with
respect to the laser field is shown on Fig.~\ref{fig:fig9}.
It is possible to understand the results of Fig.~\ref{fig:fig9} from
the symmetry of the HOMO. The $\pi_g$ symmetry means that the HOMO
has nodes along the internuclear axis as well as perpendicular to
it, and this leads to a vanishing harmonic generation when $\beta
=0^{\circ}$ and $\beta=90^{\circ}$. Consequently the harmonics take
on maximal values for intermediate angels. All harmonics are most
intense at orientations near $45^{\circ}$. It is also worth noticing 
that the precise location of the individual peak depends slightly 
on the order of the harmonic. This reflects that higher harmonics probe 
regions closer to the core.

\begin{figure}[ttb]
\includegraphics[width=\columnwidth]{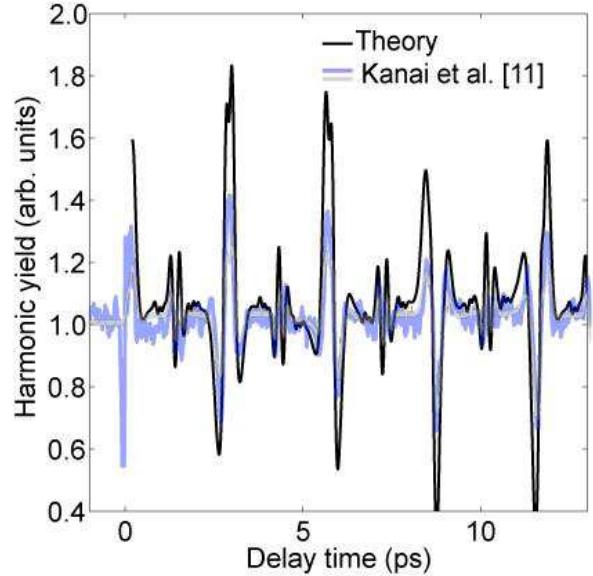}\\
\caption{(color online). Comparison of our model calculations with
the experimental results from Ref.~\cite{kanai1} in the case of
O$_2$. We have calculated the intensity of the $23$rd harmonic
from a rotational wave packet of O$_2$ normalized against the
randomly aligned case.}\label{fig:fig02}
\end{figure}

For the O$_2$ molecule, as well, several measurements of HHG from a
non-adiabatically aligned sample have been carried
out~\cite{itani2,kanai1,miyazaki}. We compare our model calculations
with the results presented in Ref.~\cite{kanai1} in the case of HHG
from O$_2$. In retrieving the time-dependent angular distribution we
use the experimental pump pulse of $50$ fs with wavelength $800$ nm
and intensity $6\times10^{13}$ W/cm$^2$. We calculate the value of
$d_N(\beta)$ for use in Eq.~\eqref{eq:dNtime} under assumptions
corresponding to the experimental probe pulse with wavelength
$\lambda=800$ nm and intensity $I=2\times10^{14}$ W/cm$^2$.
As in the case of N$_2$ above, we normalize the harmonic intensity
to the intensity of a randomly aligned sample in our theoretical
results. The comparison is shown on Fig.~\ref{fig:fig02}, and we
observe a fine agreement between theory and experiment.

\subsection{The isotopic molecules H$_2$ and D$_2$}
We now turn to a short discussion of calculations carried out for HHG 
from the isotopic molecules H$_2$ and D$_2$. Harmonic spectra from 
these molecules provide a way to study the effect of nuclear motion, 
because the FC distributions of both H$_2$ and D$_2$ are broad. 
Furthermore a comparison of the HHG signals from the isotopes contains 
information on how the nuclear masses affect the harmonic spectrum.

\begin{figure}[ttb]
\includegraphics[width=\columnwidth]{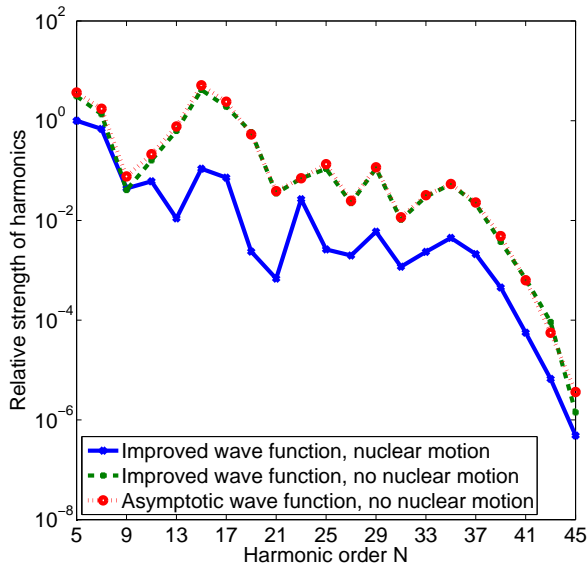}\\
\caption{(color online). HHG from the H$_2$ molecule. Calculations
are performed under the setup mentioned in the text, and all curves
are normalized to the strength of the fifth harmonic obtained by
using the improved wave function in the LAR step and taking into
account the effect of nuclear motion.}\label{fig:fig5}
\end{figure}

In Fig.~\ref{fig:fig5}, showing HHG from the H$_2$ molecule,
we have used the same laser parameters as in the case of N$_2$
and O$_2$ studied above, i.e., 800 nm, $2\times10^{14}$W/cm$^2$.
As seen from the figure, it is essential to include nuclear
motion in the case of H$_2$, and the general effect of the nuclear
motion is to reduce the intensity of harmonic generation. This can
easily be understood from the fact that the FC distributions of the
isotope is broad. The broadness implies that a substantial part of
the harmonic signal originates from transitions, where the
ionization potential is higher than in the fixed nuclei picture.
As a result the whole process gets more unlikely and so the radiative
emission is less intense. Very similar observations and conclusions
hold true in the case of D$_2$.

\begin{figure}[ttb]
\includegraphics[width=\columnwidth]{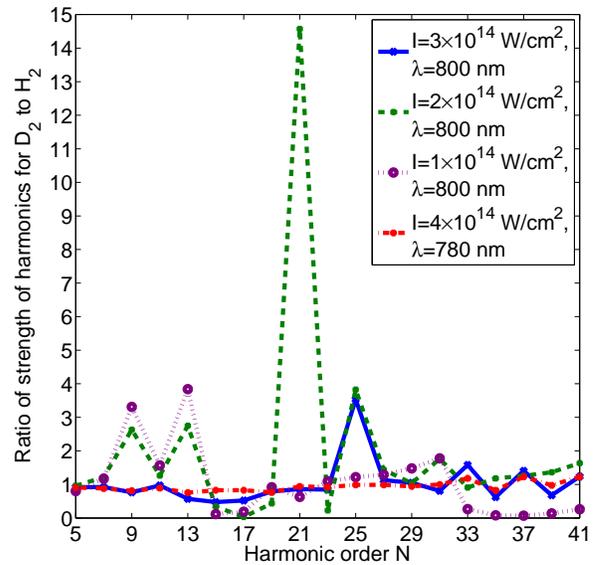}\\
\caption{(color online). Comparison of harmonic generation for the
isotopes H$_2$ and D$_2$ for different intensities at orientation
$\beta=90^{\circ}$. The ordinate shows the strength of the $N$th 
harmonic of D$_2$ divided by the strength of the $N$th harmonic of 
H$_2$. The laser parameters for the dash-dotted curve are the same 
as those used in Ref. \cite{lein1}.}\label{fig:fig7}
\end{figure}

To see how the nuclear mass influences HHG, we compare the
harmonic generation of the two isotopes for different intensities at
a laser energy of $\omega = 0.057$ a.u. in a geometry, where the
internuclear axes of the molecules are aligned perpendicular to the
laser polarization. In order to be able to compare with results of
Ref.~\cite{lein1} we have furthermore included a calculation at
intensity $I=4\times10^{14}$ W/cm$^2$ at energy $\omega=0.0584$
a.u., although our assumption of negligible depletion of the ground
states is questionable in this situation. The results are
shown in Fig.~\ref{fig:fig7}.
In Ref.~\cite{lein1} a two-dimensional model was used and it
was found that the heavier isotope leads to stronger harmonic generation. 
We cannot conclude something as simple from Fig.~\ref{fig:fig7}. Our 
non-conclusive picture of the relation between nuclear mass and HHG 
is supported by others \cite{mccann}, and is related to the strong 
correlation of the vibrational states of the residual molecular ion to 
the momentum of the excited electron as follows from Eq.~\eqref{dN}.

\section{Summary and conclusion}\label{Conclusion}
We have developed the strong-field approximation for HHG in diatomic
molecules for calculating the harmonic signal from a single diatomic
molecule of arbitrary orientation with respect to a monochromatic
linearly polarized laser field. The model incorporates the effect of
nuclear motion.

We calculated the high-harmonic emission from the molecules N$_2$,
O$_2$, H$_2$ and D$_2$. The two latter species were considered in
order to study the effect of nuclear motion in detail. The effect of
nuclear motion on these light molecules was significant and
generally reduced the intensity of the harmonics. However, there
seems to be no simple common relation between nuclear mass and the
intensity of the harmonic generation.

Turning to the N$_2$, O$_2$ cases, we successfully employed the
present model to reproduce experimental results on HHG from
non-adiabatically aligned N$_2$ and O$_2$
molecules~\cite{kanai1,itani2}. The spectra for N$_2$ were very
sensitive to the quality of the wave functions used to describe the
molecule, while the effect of nuclear motion was small. The
intensity of the harmonics was generally weaker, when the angle
between the laser polarization and the internuclear axis was
$90^{\circ}$ as compared to the intensity at an angle of $0^{\circ}$.
This orientation dependence characterizes diatomic molecules with a
HOMO of $\sigma_g$ symmetry. We also studied the orientation
dependence of the harmonics for the O$_2$ molecule, where the HOMO
has $\pi_g$ symmetry. As in the case of N$_2$, the quality of the
wave functions, used to model the O$_2$ molecule was of importance,
but furthermore we observed some effect of the nuclear motion on the
harmonic emission. The $\pi_g$ symmetry of the O$_2$ molecule
revealed itself through the angular dependence of the harmonics. In
the case of O$_2$ the harmonic generation was suppressed at
orientations corresponding to both perpendicular as well as parallel
alignment of the internuclear axis to the laser polarization. In
support to earlier work~\cite{zhou1}, our results show that it is
mainly the symmetry of the HOMO that determines the alignment
dependence of the HHG.

To conclude, we have presented a length gauge strong-field
approximation for HHG in diatomic molecules. The theory is very
versatile and readily adapted to a wide range of diatomic molecules,
and straightforwardly generalized to polyatomic molecules. The
agreement with recent pump-probe experiments~\cite{kanai1,itani2}
is excellent, and hence we have confidence in the predictive power
of the model. Despite the good agreement, it is clear that the
model poses important questions, in particular concerning the fact
that our model is formulated in the length gauge. This seems to be
advantageous for the initial ionization step in order to obtain the
correct orientation dependence in the process, and it is also physically
well-justified since the ionization is tunneling like and the
escape into the continuum happens at large distances, where the length
gauge interaction has large weight~\cite{kjeldsen1}. The physics in
the recombination step, however, is in a sense completely the opposite.
In this case an effective transfer of momentum has to take place between the
recombining electron and the core. Such a process is expected to be
better described in the velocity gauge which naturally probes
regions in space where the electron experiences abrupt changes
(close to the nuclei).

Another point is the question of how to treat the nuclei. The
present formalism gives an attempt in that direction:
within the Born-Oppenheimer approximation, we applied the
Franck-Condon principle and observed significant
effects of the nuclear motion only for the lighter species H$_2$, and
D$_2$. If it holds true then nuclear motion is essentially unimportant
for all heavier systems, and it means that
we may think of the nuclei as stationary fixed-in-space objects setting
up an effective diffraction-like grating for the backscattered electron.
Of course the out-come of the rescattering be it HHG or, e.g., future
ionization, will be highly sensitive to this spatial arrangement, and
this will undoubtedly lead to a higher degree of control of the
harmonics.

\begin{acknowledgments}
We would like to thank T. K. Kjeldsen and C. Z. Bisgaard for their
contributions to the results presented in Sec.~\ref{sectionResults}.
L.B.M. is supported by the Danish Natural Science Research Council
(Grant No. 21-03-0163) and the Danish Research Agency (Grant. No.
2117-05-0081).
\end{acknowledgments}

\appendix

\section{The asymptotic molecular wave function}
\label{AppendixA} In this appendix we work out the asymptotic form
of the momentum space wave function of the HOMO and describe, how this
expression reduces to a well known function around
singular points~\cite{kuchiev1}. We start out in the body fixed
frame of the molecule with the $z$ axis chosen along the
internuclear axis. The HOMO wave function in position representation
must then follow the asymptotic Coulomb form
\begin{equation} \label{asym-space-wave}
 \phi_0^{bf}(\mathbf{r})\sim \sum_{l} C_{l,m} r^{Z / \kappa -1}\exp(-\kappa r)
  Y_l^m(\hat{\mathbf{r}}),
\end{equation}
where $\hat{\mathbf{r}}$ is a unit vector in direction of
$\mathbf{r}$, $-\kappa^2/2 = E_b$ is the binding energy of
the HOMO-electron, and $Z$ is the net charge of the molecule, when 
the HOMO electron has been removed. $Y_l^m(\hat{\mathbf{r}})$ is a 
spherical harmonic. The summation is over the set of $l$'s such that 
$C_{l,m} \neq 0$. In this article we are concerned with terms corresponding
to $l=0,2,4$ (c.f. Table~\ref{tab:table1}).

The Fourier transform of Eq.~\eqref{asym-space-wave} yields the momentum
space wave function
\begin{widetext}
\begin{eqnarray}\label{asym-momentum-wave}
  \tilde{\phi}_0^{bf}(\mathbf{q}) &\sim& 
  \sum_{l}\sqrt{2}C_{lm}
  \left(\frac{q}{i\kappa}\right)^{l}\frac{1}{2^{l+1}\kappa^{2+Z / \kappa}}
  \frac{\Gamma (l+2+Z / \kappa)}{\Gamma(l+\frac{3}{2})}Y_l^m(\hat{\mathbf{q}})\nonumber \\
  &\times&_2F_1\left(\frac{l+2+Z / \kappa}{2},\frac{l+2+Z / \kappa}{2} + \frac{1}{2}; l+
  \frac{3}{2}; -\left(\frac{q}{\kappa}\right)^2\right),
\end{eqnarray}
\end{widetext}
where $\hat{\mathbf{q}}$ is a unit vector in the direction of
$\mathbf{q}$. By evaluating Gauss' hyper geometric functions of type
$_2F_1\left(a,a+1/2;l+3/2;z^2\right)$ and then insert $a =
(l+2+Z / \kappa)/2$ and $z^2 = - \left(q/\kappa\right)^2$ we can obtain an
explicit representation of the Fourier transform in
Eq.~\eqref{asym-momentum-wave}. To determine an expression of Gauss'
hyper geometric function we have used formulas (15.1.9) and
(15.1.10) in Ref.~\cite{abramowitz} along with a recursion relation
coming from (9.137,1) in Ref.~\cite{gradsthein1}. Consequently, we
obtained expressions of type
\begin{widetext}
\begin{eqnarray}
  _2F_1\left(a, a + \frac{1}{2};\frac{2n+1}{2};z^2 \right)= (-1)^n
  \frac{(2n-1)(2n-3) \ldots 3}{2}\left(1-z^2\right)^{n-2a} z^{-n}
  \nonumber \\
  \times (2a-(2n-1))^{-1} (2a-(2n-2))^{-1}(2a-(2n-3))^{-1}
  \ldots (2a-1)^{-1} \nonumber \\
  \times \left[\frac{\mathscr{P}(+,2(n-1))}{\left[z(1+z)\right]
  ^{n-1}}(1-z)^{2a-n}-\frac{\mathscr{P}(-,2(n-1))}{\left[z(1-z)
  \right]^{n-1}}(1+z)^{2a-n}\right],\label{generalExpression}
\end{eqnarray}
\end{widetext}
where $\mathscr{P}(\pm,2(n-1))$ is a polynomial of degree $2(n-1)$.
The first five of these polynomials are given in Table
\ref{tab:table2}.
\begin{table*}
\caption{\label{tab:table2}Polynomials of
Eq.~\eqref{generalExpression} for use in the asymptotic form of the
HOMO momentum space wave functions.}
\begin{ruledtabular}
\begin{tabular}{cl}
 $n$&$\mathscr{P}(\pm,2(n-1))$\\ \hline
 $1$&$1$ \\
 $2$&$1\pm(2a-1)z+(2a-2)z^2$\\
 $3$& $3\pm3(2a-1)z+(4a^2-7)z^2\pm(8a^2-18a+7)z^3+(4a^2-12a+8)z^4$\\
 $4$& $15\pm15(2a-1)z+6(4a^2-a-8)z^2\pm2(4a^3+12a^2-55a+24)z^3+(24a^3-72a^2+6a+57)z^4$\\
 & $\pm(24a^3-120a^2+168a-57)z^5+8(a^3-6a^2+11a-6)z^6$\\
 $5$& $105\pm(210a-105)z+15(12a^2-4a-29)z^2 \pm 5(16a^3+24a^2-190a+87)z^3+(16a^4+160a^3-760a^2+200a+699)z^4$ \\
 & $\pm (64a^4-160a^3-640a^2+1750a-699)z^5+(96a^4-640a^3+1140a^2-140a-561)z^6$\\
 & $\pm(64a^4-560a^3+1640a^2-1810a+561)z^7+16(a^4-10a^3+35a^2-50a+24)z^8$\\
\end{tabular}
\end{ruledtabular}
\end{table*}

We are also interested in the limit $q\to\pm i\kappa$ corresponding to
singularities of the asymptotic wave function, because this
expression is used for the HOMO wave function in Sec.~\ref{Theory},
when the $t'$ integration is evaluated by the saddle-point method.
The continuation of Eq.~\eqref{asym-momentum-wave} to the complex $\mathbf{q}$ 
plane is straight-forward \cite{becker}. Around the points of singularity 
Eq.~\eqref{asym-momentum-wave} can be written as~\cite{kuchiev1}
\begin{widetext}
\begin{equation}\label{singularity-wave}
  \tilde{\phi}_0^{bf}(\mathbf{q}) \sim \sum_{l}
  \left(\frac{2}{\pi}\right)^{1/2} C_{l,m}(\pm1)^l
  Y_l^m(\hat{\mathbf{q}})\frac{(2\kappa)^{Z / \kappa}\Gamma(1+Z / \kappa)}
  {(q^2+\kappa^2)^{1+Z / \kappa}},
\end{equation}
\end{widetext}
where $\pm 1$ corresponds to $q\to\pm i\kappa$. For the $l$-values
used in this work Eq.~\eqref{singularity-wave} can be verified
directly by calculating limits of the explicit representation of the
relevant Gauss' hyper geometric functions from
Eq.~\eqref{generalExpression} and substituting these expressions
into the terms in Eq.~\eqref{asym-momentum-wave}.


\end{document}